# Density functional investigation of spin polarization in bulk and thin films of nitrogen intercalated $Cu_3N$

Seyed Mojtaba Rezaei Sani[1], Masoud Karimipour[2,*], Marzieh Ghoohestani[3], Seyed Javad Hashemifar[4]

[1] Computational Physical Sciences Research Laboratory, School of Nano-Science, Institute for Research in Fundamental Sciences (IPM), Tehran, Iran

[2] Department of Physics, Vali-E-Asr University of Rafsanjan, 77139-36417 Rafsanjan, Iran

[3] Electroceram research center, Department of Physics, Malek Ashtar University of Technology, Shahin Shahr, Isfahan, Iran

[4] Department of Physics, Isfahan University of Technology, Isfahan, 84156-83111, Iran

**\* Corresponding author, E-mail address: masoud.karimipour@gmail.com**

Tel: +98-935-226-2987     Fax: +98-391-320-2024

Abstract:

It has been reported theoretically that the intercalation of nitrogen in the voids of the rather open cubic structure of bulk $Cu_3N$ build up a magnetic structure. In an extended effort to study this system, we have investigated spin polarization in bulk and thin films of nitrogen intercalated $Cu_3N$ ($Cu_3N_2$) structure by means of first-principles calculations based on Kohn-Sham density functional theory and ultrasoft pseudopotentials technique. Contrary to the previous study, the results show that after an accurate structural relaxation of the system, magnetism in the bulk structure vanishes. This effect is due to the migration of the intercalated nitrogen atom from the body center of the cell to the nearness of one of the cell faces. Similar study for the thin films of 5, 7, 9 and 11 monolayers thickness was performed and it was found that initial relaxation of structures with 7 and 11 monolayers show a net magnetic moment of 2.6 $\mu_B$. By a more extended survey of the energy surfaces, the film with 7 monolayers loses its magnetic moment similar to the bulk structure but the film with 11 monolayers maintains its magnetic moment. It is possibly a new quantum size effect that keeps the intercalated nitrogen atom of the middlemost cell at the body center site. Electron density map of this film clearly confirms the spin polarization upon the intercalated atom.

Introduction

In the search of weak itinerant magnetic materials composed of usually non-magnetic elements, several atoms have been inserted to $Cu_3N$ theoretically [1-5] and experimentally [6-10]. $Cu_3N$ has an anti-$ReO_3$ type structure (space group: Pm3m, experimental lattice constant: 3.819 Å [11,12]) with an empty atom position in the body center of its cubic close-packed cell. Atom intercalation into this empty body center site and its effects on physical properties of the system has motivated several studies. For example First-principle studies on $Cu_3NM$ (M = Ni, Cu, Zn, Pd, Ag, Cd, Rh, and Ru) revealed that the addition of such metal atoms modifies the electronic properties of $Cu_3N$ and turns all copper-metal-nitrides properties into the metallic ones [1,2]. Especially, RhCu$_3$N could be a good candidate for exhibiting either quantum critical behavior or itinerant electron meta-magnetism [1].

N intercalated Cu$_3$N, Cu$_3$N$_2$, has been also studied by several groups [3,6,13-15]. Hou [3] has shown theoretically that nitrogen intercalation into Cu$_3$N structure results in semiconductor to metal phase transition and this structure could not be mechanically stable. Experimental investigations [15] confirm that in the N-rich Cu$_3$N structure, nitrogen migrates from bulk to the film surface, without out-diffusing. The XRD data along with Rietveld refinement are suggesting that the excess nitrogen atoms possibly accommodate in the body center interstitial sites of the anti-ReO$_3$ crystal lattice and form solid solution. They also observed [13] direct optical transitions corresponding to the stoichiometric semiconductor Cu$_3$N plus a free-carrier contribution that can be tuned in accordance with the N-excess. However, so far, the long-standing problem of extra nitrogen location has been unanswered.

Recently, Ghoohestani et al. [5] showed that the body-centered N-intercalated Cu$_3$N$_2$ structure undergoes spin polarization. It was demonstrated that the application of the Perdew–Burke–Ernzerhof (PBE) [16] and modified Becke–Johnson (mBJ) parameterizations [17] of the generalized gradient approximation (GGA) results in magnetic behavior of the Cu$_3$N$_2$ structure and the intercalation is an endothermic process which, within PBE, give rises to a nearly half metallic behavior, while mBJ favors semiconductor magnetism. The net magnetic moment of the system was shown to be 2.3 $\mu_B$ per formula unit (f.u.) and the 2$p$ orbital of the intercalated nitrogen atom is responsible for the spin polarization of the system.

In this work, we investigate magnetic properties and stability of N intercalated bulk and thin films of $Cu_3N$ structure. To the best of our knowledge, no report has discussed the origin of the magnetization in the bulk $Cu_3N_2$. First, it is tried to address the origin of magnetism in the bulk structure of $Cu_3N_2$. Thereafter we try to study the magnetic behavior of $Cu_3N_2$ thin films to find the possible effects of surface and quantum size on the behavior of intercalated nitrogen atoms.

II. Computational method

We performed first-principles calculations based on Kohn-Sham density functional theory [18] and ultrasoft pseudopotentials technique [19] by means of Quantum ESPRESSO package [20] to investigate the electronic, magnetic, and structural properties of the target systems. The relativistic effects are taken into account in the scalar approximation, but the spin-orbit coupling is neglected which is expected to be small in light atoms like Cu and N. The exchange-correlation functional was described by the PBE-GGA [16] and for comparison by local density approximation (LDA) [21]. Absence of the correlated d or f electrons may guaranty the efficiency and reliability of this conventional GGA functional for computing the electronic and magnetic properties of our systems.

We started with the experimental lattice parameter value (3.82 Å) [12]. A kinetic energy cut off of 35 Ryd for the plane wave basis set expansion, and 300 Ryd for the Fourier expansion of electron density were applied. The Brillouin-zone integrations were performed on a regular mesh of 9×9×9 k points for bulk structure and equivalent meshes for thin film structures. The structures are accurately relaxed to achieve residual forces less than 0.01 mRyd/Bohr and energy accuracy of about 0.001 mRyd/f.u. The supercell approach with a vacuum region of 20 Bohr thick was used to simulate two dimensional thin films in the three dimensional periodic boundary conditions imposed by the code.

III. Results and discussion

The calculated electronic structure of the body-centered N-rich bulk copper nitride which we call it ideal bulk $Cu_3N_2$ (Figure 1a) shows a net magnetic moment of about 2.3 $\mu_B$, mostly results from $2p$ orbital of inserted N atom, consistent with the reported full-potential calculations [5]. We know that this structure could not be mechanically stable [3] so the N atom at the body center position was slightly displaced and then the structure was allowed to relax. It was observed that the intercalated N atom significantly moves towards the face center position (Figure 1b) and consequently the spin polarization disappears. Indeed, the body center intercalation of the system put it in a local minimum of energy with magnetic characteristics, while, displacing the inserted nitrogen bring the system to a non-magnetic global minimum. The total energy calculation shows that the relaxed bulk structure (Fig. 1b) of $Cu_3N_2$ is more stable than its ideal bulk counterpart by means of 52 mRyd. It is then inferred from the results that the spin polarization of $Cu_3N_2$ is very sensitive to the position of the intercalated nitrogen atom. In the other words, the system carries magnetic moment only when the intercalated nitrogen occupies precisely the body center site, which is an instable configuration. The free nitrogen atom has a Hund magnetic moment of $3\mu_B$, which has been reduced to 2.3 $\mu_B$ in the body center site of bulk $Cu_3N$. This small reduction could be due to the rather small crystal filed of $Cu_3N$ in its body center position.

In order to find the possible effects of confinement and surfaces on the magnetization and structure of the system, thin films of $Cu_3N_2$ with 5, 7, 9 and 11 monolayers were simulated in appropriate supercells (Figure 2). A primarily structural relaxation revealed that the magnetization of the systems in their ground state depends highly on the number of the $Cu_3N_2$ monolayers or equivalently the number of the body center sites in their structures. The films with 5 and 9 monolayers (2 and 4 body center sites) were non-magnetic, while the films with 7 and 11 monolayers (3 and 5 body center sites) showed spin polarization with a magnetic moment of 2.6 $\mu_B$.

To elucidate that the resultant spin polarization is not due to the employed exchange-correlation approximations, the same calculations were performed using LDA and the results confirmed the mentioned spin polarization effect. The analysis of projected density states of electronic structure of magnetic thin films clarifies that (figure 3) the polarization mostly origins in $2p$ orbital of the middlemost intercalated nitrogen atom which is symmetrically pinned and cannot move, while all other intercalated N atoms of the magnetic films as well as all intercalated N atoms of the non-magnetic films move toward the neighboring face center sites. From the figure it can be seen that $3d$ orbital of Cu atoms make a little contribution to the magnetism of the structure. Hence, similar to the bulk structure, magnetization of $Cu_3N_2$ thin films is very sensitive to the position of the intercalated nitrogen atoms.

However, the mechanical stability of these magnetic thin films is in question. To address this question, the magnetic N atom was slightly displaced from its ideal position and then the system was allowed to relax. It was observed that (Figure 4) in the 7 monolayers film, the central nitrogen atom like other inserted atoms moves towards the neighboring face center site and the film loses its magnetization, whereas 11 monolayers film maintains its magnetic moment by returning back the displaced nitrogen atom to the central body center position. It is possibly a quantum size effect that stabilizes the central body center site of the $Cu_3N_2$ thin film with 11 monolayers. Furthermore, to investigate the stability against changes in the cell volume we have calculated total energy of the system in different volumes which the results for 7 and 11 monolayers thin films fitted to the Murnaghan equation of state [22] have been shown in Fig. 5.

Electron density map of the middlemost unit cell of relaxed 11 monolayers thin film has been plotted in Fig. 6. The contour plot of the spin density confirms the conclusion that the spin polarization of the system results from the spin polarization of the nitrogen p states. From lateral view (Fig. 6 left) a slight difference between spin up and down electron density of intercalated N can be seen. The top view (Fig. 6 right) shows a more expanded electron cloud for spin up, while the spin down electron cloud is rather localized.

As an indicator of the stability of the system, the cohesive energy ($E_{coh}$) has been calculated for the relaxed thin films using the following equation:

$$E_{coh} = (E_{tot} - (n^{Cu}E^{Cu} + n^{N}E^{N}))/(n^{Cu} + n^{N})$$

Where $E_{tot}$, $E^{Cu}$, $E^{N}$ are the total energy of the structure, the energy of isolated Cu and N atoms, $n^{Cu}$ and $n^{N}$ are the number of Cu and N atoms in the supercell, respectively. The obtained cohesive energies for 5, 7, 9 and 11 monolayers thin films and the bulk structure were -4.6377, -4.6191, -4.6074, -4.5390 and -4.4170 eV/atom, respectively. It can be seen that 11 monolayers film is the most stable and the bulk structure is the less stable one.

In a further investigation, we intercalated three nitrogen atoms in one monolayer of $Cu_3N$. One of them was positioned at the center of the cell and the two others at the opposite faces of the cell which are located at the surfaces of the monolayer (Fig. 7a). After structural relaxation, the system was observed to be non-magnetic. Although in the relaxed structure of this system (Fig. 7b), the intercalated N at the body center site keeps its position, the other two intercalated N atoms move toward the central N atom to make strong covalent bond with it. This strong covalent bonding diminishes the magnetic moment of intercalated N atoms and give rise to a non-magnetic ground state for $Cu_3N_4$.

Conclusion:

To address the long-standing problem of extra nitrogen location in $Cu_3N_2$ we have performed simulations based on first-principles pseudopotential methods. It was argued that magnetism in the N intercalated bulk $Cu_3N$ and $Cu_3N$ thin films is very sensitive to the number and the position of the intercalated atoms. The intercalated N atom in the body center site of the cubic cell of $Cu_3N$, exhibits an atomic like magnetism which is somewhat weakened by the crystal filed of the system. In the bulk $Cu_3N$ and $Cu_3N$ thin films with 5, 7, 9 monolayers, the intercalated N atoms in the body center sites prefer to move towards the neighboring face center sites and lose their magnetization. While, in the 11 monolayers thin film, the middlemost body center site seems to be a stable position for the N intercalation and hence this system may exhibit stable magnetism. Finally, it was argued that more N enrichment in $Cu_3N$ ($Cu_3N_4$) may give rise to covalent bonding between the intercalated atoms and consequently a non-magnetic ground state.

References

[1] M. Sieberer, S. Khmelevskyi, P. Mohn, "*Magnetic instability within the series TCu$_3$N (T=Pd, Rh, and Ru): A first-principles study*", Phys. Rev. B. 74, 014416 (2006).

[2] M. M. Armenta, W. L. Perez, N. Takeuchi, "*First- principles calculations of the structural and electronic properties of Cu$_3$MN compounds with M= Ni, Cu, Zn, Pd, Ag, and Cd*", Solid State Sciences. 9, 166-172 (2007).

[3] Z. F. Hou, "*Effect of Cu, N, and Li intercalation on the structural stability and electronic structure of cubic Cu3N*", Solid State Sciences 10, 1651–1657 (2008).

[4] X.Y. Cui, A. Soon, A. E. Phillips , R. K. Zheng, Z. W. Liu , B. Delley, S. P. Ringer , C. Stampfl, "*First principles study of 3d transition metal doped Cu$_3$N*", Journal of Magnetism and Magnetic Materials, 324, 3138–3143 (2012).

[5] M. Ghoohestani, M. Karimipour, H. A. Badehian, S. J. Hashemifar, "First-principle calculation of spin polarization in Cu$_3$N$_2$", Journal of Magnetism and Magnetic Materials, 344, 202–206 (2013).

[6] N. Gordillo, R. Gonzalez-Arrabal, M. S. Martin-Gonzalez, J. Olivares, A. Rivera, F. Briones, F. Agulló -Lopez, D. O. Boerma, "*DC triode sputtering deposition and characterization of N-rich copper nitride thin films: role of chemical composition*", Journal of Crystal Growth 310, 4362–4367 (2008).

[7] A. Rahmati, "*Reactive DC magnetron sputter deposited Ti–Cu–N nano-composite thin films at nitrogen ambient*", Vacuum, 85, 853–860 (2011).

[8] L. Gao, A. L. Ji, W. B. Zhang, Z. X. Cao, "Insertion of Zn atoms into Cu$_3$N lattice: Structural distortion and modification of electronic properties", Journal of Crystal Growth, 321, 157–161 (2011).

[9] N. Gordilloa, R. Gonzalez-Arrabalb, P. Diaz-Chaoc, J. R. Aresc, I. J. Ferrerc, F. Ynduraind, F. Agulló-López, *"Electronic structure of copper nitrides as a function of nitrogen content"*, Thin Solid Films, 531, 588-591, (2013).

[10] A. L. Ji, N. P. Lu, L. Gao, W. B. Zhang, L. G. Liao, and Z. X. Cao, *"Electrical properties and thermal stability of Pd-doped copper nitride films"*, J. Appl. Phys. 113, 043705 (2013).

[11] R. Juza and H. Hahn, *"Über die Kristallstrukturen von Cu3N, GaN und InN Metallamide und Metallnitride"*, Z. Anorg. Allg. Chem. 239, 282-287 (1938).

[12] U. Zachwieja and H. Jacobs, *"Ammonothermalsynthese von kupfernitrid, Cu3N"*, J. Less-Common Met. 161, 175 (1990).

[13] N. Gordillo, R. Gonzalez-Arrabal, A. Alvarez-Herrero, and F. Agullo-Lopez, *"Free-carrier contribution to the optical response of N-rich $Cu_3N$ thin films"*, J. Phys. D: Appl. Phys. 42 165101 (2009).

[14] N. Gordillo1, A. Rivera, R. Grötzschel, F. Munnik, D. Güttler, M. L. Crespillo, F. Agulló-López, and R. Gonzalez-Arrabal, *"Compositional, structural and morphological modifications of N-rich $Cu_3N$ films induced by irradiation with Cu ions at 42 MeV"*, Journal of Physics D: Applied Physics 43, 34 345301 (2010).

[15] R. A. Gonzalez, N. Gordillo, M. S. M. Gonzalez, R. R. Bustos, F. A. Lopez, *"Thermal stability of copper nitride thin films: The role of nitrogen migration"*, J. Appl. Phys. 107, 103513 (2010).

[16] J. P. Perdew, K. Burke, M. Ernzerhof, *"Generalized gradient approximation made simple"*, Phys. Rev. Lett. 77, 3865 (1996).

[17] F. Tran, and P. Blaha, *"Accurate Band Gaps of Semiconductors and Insulators with a Semilocal Exchange- Correlation Potential"*, Phys. Rev. Lett. 102, 226401 (2009).

[18] W. Kohn and L. J. Sham, *"Self-Consistent Equations Including Exchange and Correlation Effects"*, Phys. Rev. 140, A1133 (1965).

[19] D. Vanderbilt, *"Soft self-consistent pseudopotentials in a generalized eigenvalue formalism"*, Phys. Rev. B, 41, 7892(R) (1990).

[20] P. Giannozzi, et al, *"QUANTUM ESPRESSO: a modular and open-source software project for quantum simulations of materials"*, J.Phys.:Condens.Matter, 21, 395502 (2009).

[21] J. P. Perdew and Y. Wang, *"Accurate and simple analytic representation of the electron-gas correlation energy"*, Phys. Rev. B, 45, 13244 (1992).

[22] F. D. Murnaghan, *"The compressibility of media under extreme pressures"*, Proc. Nat. Acad. Sci. 30, 244 (1944).
Figure Captions:

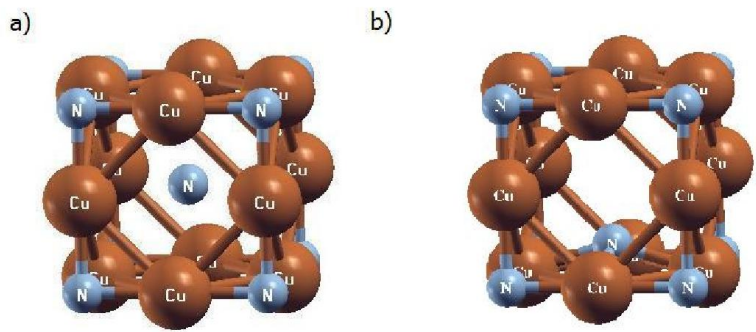

Figure 1: (a) The unit cell of ideal N-intercalated $Cu_3N$ bulk structure & (b) the relaxed unit cell of $Cu_3N_2$ bulk structure.

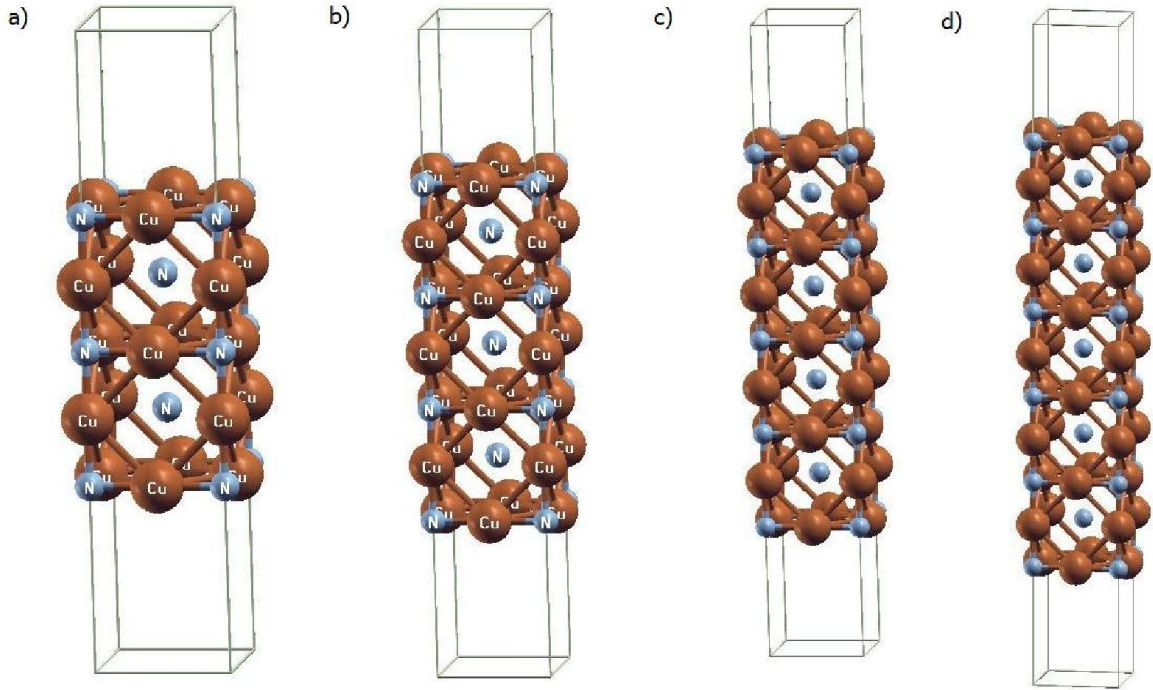

Figure 2: Considered (initial) supercells of $Cu_3N_2$ thin films with (a) 5 (b) 7 (c) 9 & (d) 11 monolayers.

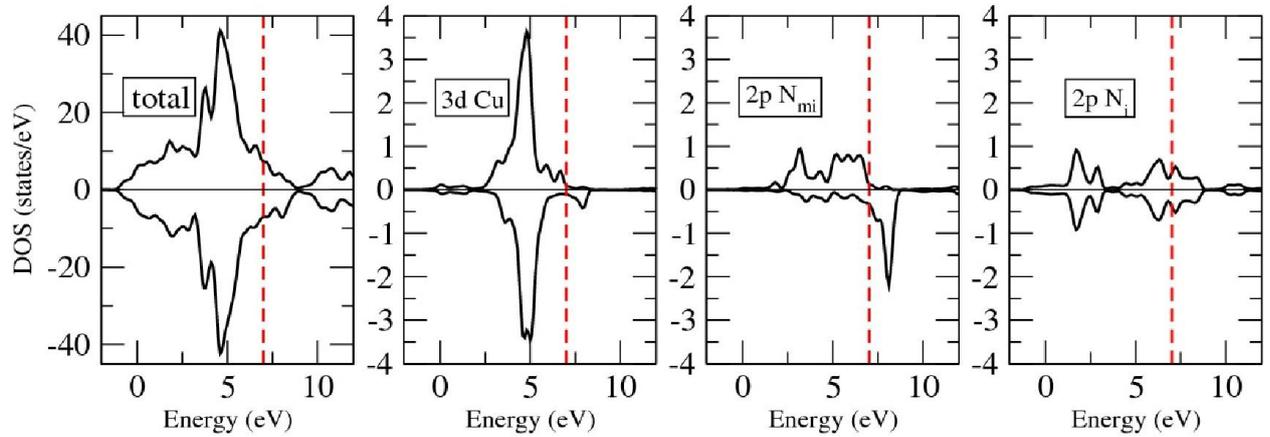

Figure 3: Total density of electronic states for 11 monolayers film as well as the projected DOS for $3d$ orbital of Cu and $2p$ orbital of middlemost intercalated N ($N_{mi}$) and one of the other intercalated N ($N_i$). Dashed lines show the Fermi energy (7.01 eV). Positive and negative values hold for spin up and spin down states, respectively.

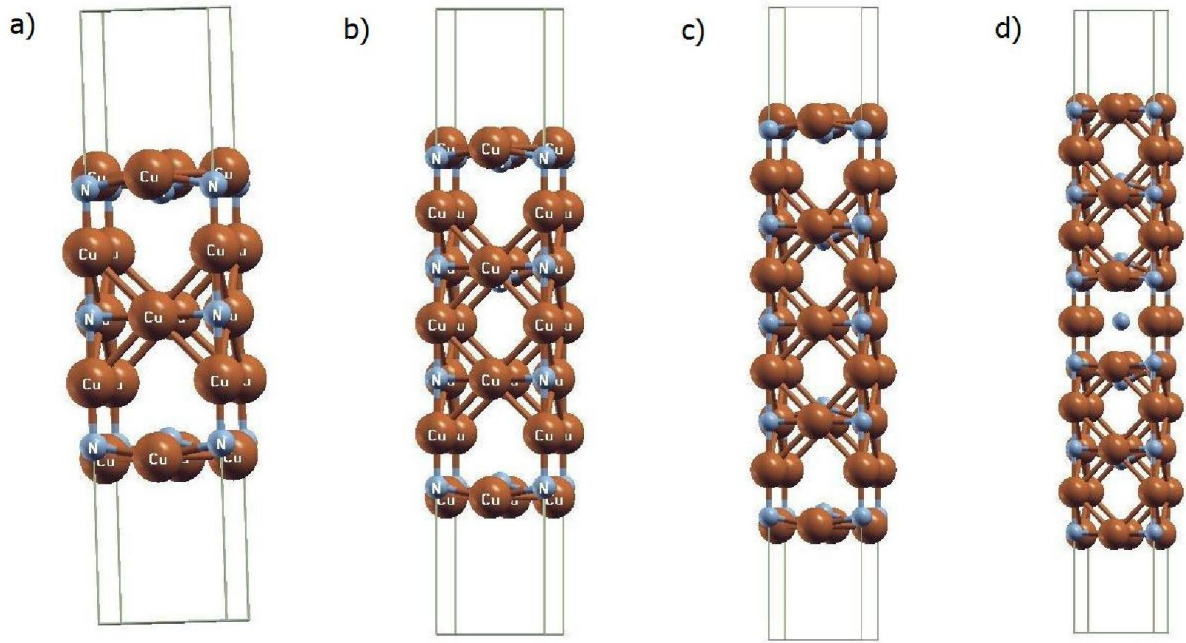

Figure 4: Relaxed structures of $Cu_3N_2$ thin films with (a) 5 (b) 7 (c) 9 & (d) 11 monolayers thickness.

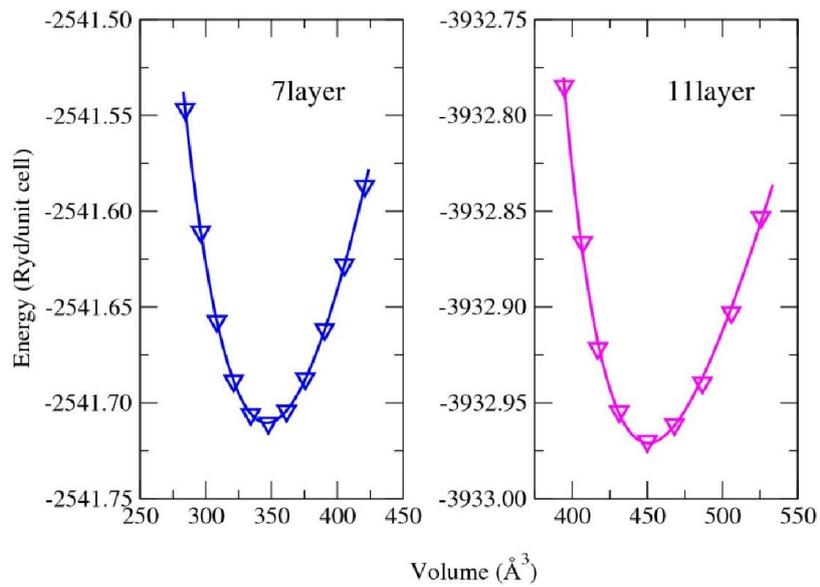

Figure 5: Total energy vs. volume for 7 and 11 monolayers thin films. The spin polarization for 11 monolayers film has been obtained at the ground state.

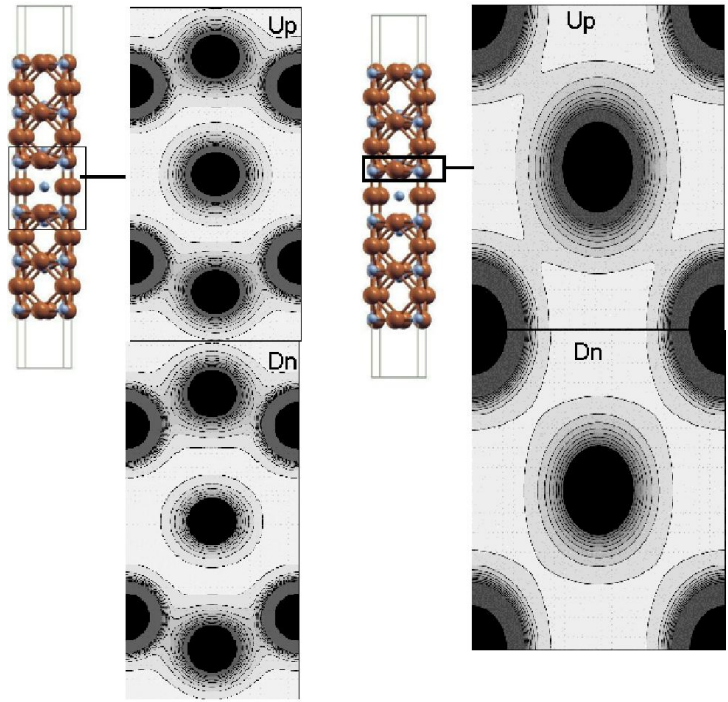

Figure 6: Electron density map of the middlemost unit cell of 11 monolayers thin film for spin up and spin down. It shows the lateral (left) and top view (right) of the unit cell.

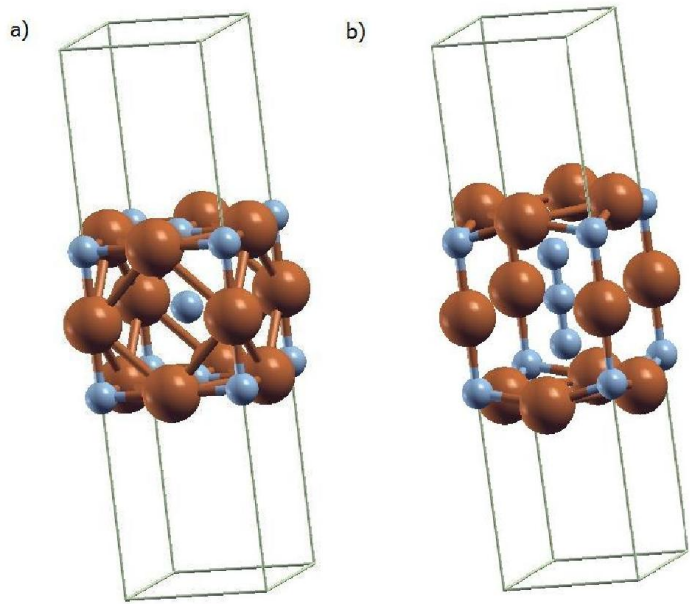

Figure 7: (a) Considered & (b) relaxed monolayer of $Cu_3N_2$ structure with insertion of two nitrogen atoms in the center of cube faces which are at the surface of the film.